\documentclass[twocolumn,prl,showpacs]{revtex4}
\usepackage{amstext}
\usepackage{amsmath}            %serve per le subequazioni
\usepackage{amssymb}            %serve per il simbolo "marchio registrato", \circledR
\usepackage{graphicx}           %serve per le figure eps, ps etcyy
\usepackage{latexsym}

\newcommand{\ket}[1]{|#1\rangle}
\newcommand{\bra}[1]{\langle #1|}
\newcommand{\bk}[1]{\langle #1 \rangle}

\renewcommand{\Re}{\mbox{Re}}

\begin{document}
\title{Spin-1/2 geometric phase driven by decohering quantum fields}
\author{ A. Carollo$^{\diamond}$$^\ddag$, I. Fuentes-Guridi$^{\dagger}$,
M. Fran\c{c}a Santos$^{\diamond}$ and  V. Vedral$^{\diamond}$}
\address{$^{\diamond}$Optics Section, The Blackett Laboratory,
Imperial College, London SW7 2BZ, United Kingdom \\
$^{\dagger}$ Perimeter Institute, 35 King Street North Waterloo, Ontario Canada N2J 2W9,\\
$^\ddag$INFM, unit\`{a} di ricerca di Milano, via Bramante 65, I-26013, Crema (CR), Italy}

\begin{abstract}
We calculate the geometric phase of a spin-1/2 system driven by a
one and two mode quantum field subject to decoherence. Using the
quantum jump approach, we show that the corrections to the phase
in the no-jump trajectory are different when considering an
adiabatic and non-adiabatic evolution. We discuss the implications
of our results from both the fundamental as well as quantum
computational perspective.

\end{abstract}

   %actually, we didn't put the pacs number in the text submitted.
                                    %These are the pacs number that we have specified in the submission form.
 \maketitle

In quantum mechanics physical states are equivalent up to a global phase which in general does not contain useful
information about the described system, and thus, can be ignored. However, Berry \cite{berry84a} surprisingly
showed that these phases can have a component of geometric origin with important observable consequences, being
the most cited examples the Aharonov-Bohm effect \cite{aharanov59a} and the spin-1/2 particle driven by a rotating
magnetic field \cite{berry84a}. These components which are gauge invariant and only depend on the path followed by
the system during its evolution, have been investigated and tested in a variety of settings and have been
generalized in several directions \cite{shapere89a}. Geometric phases are interesting both from a fundamental
point of view and for their applications, among which geometric quantum computation
\cite{ekert00a,falci,pachos00a} is one of the most important. In fact, the use of geometric phases in the
implementation of fault-tolerant quantum gates has motivated their study under more realistic situations
\cite{sioqvist00,ericsson02}. For example, when a system interacts with an environment, its quantum superpositions
may decay into statistical mixtures~\cite{zurek81a,caldeiras85} and this effect, called decoherence, is the most
important limiting factor for quantum computation.

Previous works investigate the behavior of geometric phases under some typical errors sources like random
classical fluctuations to the driving fields, as well as generic reservoirs acting in spin 1/2
evolutions~\cite{gabriele,carollo03}. All of them consider the driving field as a classical system. However, any
driving field is also a quantized system and, whenever this quantum behavior is relevant, which is the case in
many experimental situations, decoherence on these fields may play an important role. In fact, it may even become
critical, particularly when geometric phases are used to implement quantum protocols, like communication and
computational ones.

In this letter, we investigate the behavior of the geometric phase of a spin 1/2 particle interacting with a
driving magnetic field when this field is not only quantized but also subjected to decoherence. We calculate and
analyze the effect of decoherence of the driving field on both adiabatic and non-adiabatic evolutions of the joint
spin and quantized modes system. First we briefly describe the general framework of geometric phases in open
systems, developed in~\cite{carollo03}. Then we calculate Berry's phases for different interactions of spin 1/2
systems and decohering fields both in the adiabatic and non-adiabatic scenarios. Finally we point out the
differences between these two situations and how this noise source compares to previously analyzed ones.

%Recently, the phase acquired by a spin-1/2 particle adiabatically
%evolving under the presence of a quantized magnetic field, has
%been investigated \cite{fuentes-guridi02a}. Berry's phase was
%previously considered only under the evolution of classical fields
%and thus, one could not think of field decoherence. This model
%allows us to consider photon leakage in the geometric evolution of
%a spin-1/2 particle and quantum magnetic field. We use the quantum
%jump approach proposed in \cite{carollo03} where a general
%framework for the geometric phase is introduced for open systems.

Let us first consider a system described by the density operator $\rho$ and a Hamiltonian $H$. The decoherence
process due to the interaction with an environment (under the Markovian approximation) is described by the
following master equation ($\hbar=1$):
\begin{equation}
  \label{eq:mastereq}
  \dot{\rho}=\frac{1}{i}[H,\rho]-\frac{1}{2}\sum_{k=1}^{n}
  \{\Gamma_k^\dagger \Gamma_k\rho+\rho \Gamma_k^\dagger \Gamma_k\ - 2\Gamma_k
  \rho\Gamma_k^\dagger\},
\end{equation}
where the commutator generates the coherent part of the evolution and the remaining part represents the effect of
the reservoir on the dynamics of the system. The action of each $\Gamma_k$ amounts to a different decohering
process. Suppose that we monitor the system and do not detect any decay. The geometric phase for the ''no-jump"
trajectory for master equation (\ref{eq:mastereq}) in the continuous limit is given by \cite{carollo03}:
\begin{equation}\label{eq:gpnojump2}
  \gamma^0=\int_0^T \frac{\bra{\psi^0(t)}H\ket{\psi^0(t)}}{\bk{\psi^0(t)|\psi^0(t)}} dt -
\arg\{\bk{\psi^0(T)|\psi^0(0)}\},
\end{equation}
where
\begin{equation}
\label{eq:nojumpcont} i\frac{d}{dt}
\ket{\psi^{0}(t)}=\tilde{H}\ket{\psi^{0}(t)},\quad
\ket{\psi^0(0)}=\ket{\psi_0},
\end{equation}
and $\tilde{H}$ is a non-Hermitian effective Hamiltonian given by:
\begin{equation}
  \label{eq:nonHerHam}
  \tilde{H}=H-\frac{i}{2}\sum_{k=1}^n \Gamma_k^\dagger \Gamma_k.
\end{equation}
Before applying this general framework to our problem we consider the adiabatic
phase of a two-level system and a
single mode quantum field following \cite{fuentes-guridi02a}. We describe the
two-level system with Bohr frequency
$\omega$ in terms of Pauli operators $\sigma_z$, $\sigma_{\pm}=(\sigma_{x}\pm i\sigma_{y})/2$ and the field with
frequency $\nu$ in terms of the creation and annihilation operators $a$ and $a^{\dagger}$. In the interaction
picture, the initial Hamiltonian reads
\begin{equation} \label{jaynes}
  H_{int}=\frac{\Delta}{2}\sigma_z +g(a^{\dagger}\sigma_{-}+a\sigma_{+}),
\end{equation}
where $\Delta=\omega-\nu$ is the detuning between the quantum mode and the two-level system and $g$ is the
coupling constant. The evolution of the system is dictated by the usual time dependent Schr\"{o}dinger equation.
Following~\cite{fuentes-guridi02a}, to generate a geometric phase the Hamiltonian is varied in a cyclic and
adiabatic fashion by means of the unitary operation $U=e^{-i\phi\hat{n}}$ where $\hat{n}=a^{\dagger}a$ is the
number of photons in the field. Using Berry's formula for the phase \cite{berry84a} we find that the eigenstate of
(\ref{jaynes})
\begin{eqnarray} \label{eq:istate}
 |\Psi^{+}\rangle=\cos(\theta_{n}/2)|e,n\rangle+\sin(\theta_{n}/2)|g,n+1\rangle,
\end{eqnarray}
acquires the phase
\begin{equation}
\gamma_{+}=2\pi\langle n\rangle=2\pi n+\pi(1-\cos{\theta_{n}}),
\end{equation}
where $\cos\theta_{n}=\frac{\Delta}{2 R}$ with $R=\sqrt{\Delta^{2}/4+g^{2}(n+1)}$, when $\phi$ is varied from 0 to
$2\pi$. To consider now that the field is subject to decoherence we describe our system by equation
(\ref{eq:mastereq}) with $\Gamma=\sqrt{\lambda}a$, i.e. the field is linearly losing photons to its reservoir.
Thus, the effective Hamiltonian is $\tilde{H}=H-i\frac{\lambda}{2}\hat{n}$. Using equation (\ref{eq:gpnojump2})
and (\ref{eq:nojumpcont}) we obtain the phase
\begin{equation}\label{eq:adiab1}
\gamma_{+}^{d}= 2\pi
n+\pi\left(1-\Re\frac{\Delta-i\lambda/2}{\sqrt{(\Delta-i\lambda/2)^{2}+4g^{2}(n+1)}}\right).
\end{equation}
This result can be easily derived using the bi-orthogonal basis
technique illustrated in~\cite{garrison88} which gives rise to a
complex geometric phase whose real part corresponds, in the
adiabatic case, to the general formula~(\ref{eq:nojumpcont}).

Notice that the lowest order correction in $\lambda/R$ of
eq.~(\ref{eq:adiab1}) is quadratic, which means that in case of
low decoherence we recover the phase $\gamma_{+}$, up to the first
order in $\lambda/R$. To the second order in $\lambda/R$ the
geometric phase~(\ref{eq:adiab1}) reads:
\[
\gamma_{+}^{d}\simeq \gamma_{+} + \pi\cos\theta_n\frac{(\lambda/R)^2}{16}\left(1+3/2\cos^2\theta_n\right) .
\]

It is important to notice that in case of low decoherence the
deviation of the geometric phase from the value $\gamma_+$ is null
up to the first order in $\lambda/R$. This reflects the resilience
of the geometric phase against the environment. The reason for
this can be interpreted heuristically as a consequence of the
adiabatic evolution of the system under consideration.

Under the adiabatic assumption, the state of a system, in its evolution, tends to follow
 the eigenstates of the
instantaneous Hamiltonian. Loosely speaking, this behaviour
opposes to the tendency of the environment of dragging the state
away from its undisturbed evolution.

More precisely, the adiabatic approximation ensures that the probability for a state to follow the instantaneous
eigenstate of the unperturbed Hermitian Hamiltonian is highly enhanced compared to the probabilities of
transitions to other eigenspaces. In fact, the probability amplitudes associated to these transitions are averaged
down due to their high frequency evolutions (of the order of Bohr frequencies of the system), whereas the
probability amplitude of staying in the same eigenspace is almost stationary. If the decoherence rate is
sufficiently small, this amounts to effectively projecting the state back to the original eigenspace, whenever it
tends to be driven away by the environment. Therefore, the state trajectory on the projective Hilbert space tends
to be unaffected by the decoherence (up to the first order), thereby leaving the area enclosed in the path, and
hence the geometric phase, unchanged.

This characteristic of robustness against decoherence is also present in other systems. For example, similar
properties can be observed in the analogous model, analyzed in ref. \cite{fuentes-guridi02a}, in which a two-level
atom interacts with not one but two quantized modes of a harmonic oscillator. In this case, the geometric phase is
obtained by an adiabatic evolution of the initial Hamiltonian
\begin{equation}\label{eq:ham2mode}
H=\nu a^{\dagger}a+\nu b^{\dagger}b +g(\sigma_{+}a+\sigma_{-}a^{\dagger}),
\end{equation}
where the extra mode is described by the creation and annihilation operators $b$ and $b^{\dagger}$ respectively.

Analogously to the previous case, an initial eigenstate of this Hamiltonian $|\psi\rangle=\cos \theta_{n}/2
|e,n,n^\prime\rangle+\sin \theta_{n}/2|g, n+1, n^\prime \rangle$, is considered. After a cyclic evolution of this
state following the adiabatic rotation of Hamiltonian (\ref{eq:ham2mode}) in a two-dimensional parameter space,
the original state acquires a geometric phase equal to:
\begin{equation}
  \label{eq:twomodegph}
  \chi_{(n,n^{\prime})}=\frac{1}{2}\Omega\left[n-n^{\prime}+\frac{1}{2}(1-\cos{\theta_{n}})\right],
\end{equation}
where $\Omega$ is the solid angle described by the parameters. In ref.~\cite{carollo02} the authors suggest a
cavity QED experiment that implements this rotation and allows for the measurement of the above mentioned phase.

In order to study the case where the interacting fields are decohering, we consider the Hamiltonian
$\tilde{H}=H-i\frac{\lambda}{2}\hat{N}$ with $\hat{N}=a^{\dagger}a + b^{\dagger}b$ the total number of photons in
the system. As in the single-mode case, here a non-Hermitiam Hamiltonian is obtained from the assumption that no
jump occurs during the evolution, i.e. the system is assumed to be continuously monitored by detectors and no
emission of photon is registered. The adiabatic phase then yields
\begin{eqnarray}\label{eq:gp2mode}
&&\chi_{(n,n^{\prime})}^{d}=\\&&=\frac{\Omega}{2}\left((n-n^{\prime})+1-\Re\frac{\Delta-i\lambda/2}
{\sqrt{(\Delta-i\lambda/2)^{2}+4g^{2}(n+1)}}\right).\nonumber
\end{eqnarray}
Again, the expected geometric phase~(\ref{eq:twomodegph}) is
recovered in case of low decoherence, and the lowest order
correction is only quadratic in $\lambda/R$, being
$R=\sqrt{\Delta^{2}/4+g^{2}(n+1)}$ the Rabi frequency. %Note that,
%in these calculations we considered the specific case of fields
%decohering with the same rate which is reasonable for many
%experimental cases. Besides, the generalization for different
%decohering rates is immediate and does not add any important
%information to our result.
Finally, note that a second order correction in the decaying factor $\lambda$ for the ''no-jump" trajectory
suggests that the fields decoherence may not play such an important role in the realization of the proposed
experiment~\cite{carollo02}.

It is interesting to compare the geometric phase due to an
adiabatic evolution in presence of decoherence and the analogous
result obtained in a non-adiabatic fashion. It is well
known~\cite{aharonov87} that the adiabaticity is not a necessary
condition to observe geometric phases. In fact, these are uniquely
defined by the path on the projective Hilbert space traversed by
the quantum system in its evolution. Thus, no matter how this
evolution is achieved, the geometric phase will remained
unchanged. This may no longer be the case in presence of
decoherence, as the interaction with the environment can affects
differently adiabatic and non-adiabatic evolutions.

A typical example of non-adiabatic evolution is the one in which the Hamiltonian is time-independent and the
initial state is chosen to be a cyclic state, i.e. a state that after a suitable time T evolves back to itself, up
to a phase change. In some cases it is possible to choose a Hamiltonian and a cyclic state such that they
generates exactly the same path as the one associated to a given adiabatic evolution. In \cite{bose02} we
considered the non-adiabatic version of the fully quantised spin-1/2 phase. Here we present the analysis of this
case taking into account the effect of decoherence and compare the geometric phase obtained in the analogous
adiabatic case.

In the non-adiabatic setup that we are presenting, the system is
initially prepared in an entangled state of atom and the field
mode. Then it evolves under a time independent Hamiltonian
involving only the degrees of freedom of the field.
 By
turning on a Jaynes-Cummings interaction we can prepare the system
in the state (\ref{eq:istate}).

After this interaction has been switched off, we assume that the
dynamics of the system is described, in the interaction picture,
by the Hamiltonian:
\begin{equation}
  \label{eq:rotation}
  H_{int}=\beta \hat{n},
\end{equation}
where $\beta$ is a constant parameter. Thus, the state evolves
according to:
\begin{equation}
  \label{eq:evol}
  |\Psi^{+}(t)\rangle=e^{-i\beta \hat{n}t}|\Psi^{+} \rangle,
\end{equation}
and after a time $T=4\pi/\beta$ the state completes a closed loop.
Using the definition of Aharanov and Anandan geometric
phase~\cite{aharonov87} it is easy to show that after a cyclic
evolution the phase acquired by the state is $\gamma_{+}$, the
same as~(\ref{eq:adiab1}) obtained in the adiabatic case.

When the decoherence of the field is considered, the phase for the
no-jump trajectory can be calculated from the
expression~(\ref{eq:gpnojump2}) which yields
\begin{equation}\label{eq:gphnoad1}
\gamma^{d}=-\frac{\beta}{\lambda}\ln\bra{\Psi^+}
e^{-2\pi\frac{\lambda}{\beta}\hat{n}}\ket{\Psi^+}.
\end{equation}
 Given the initial
state~(\ref{eq:istate}), the eq.~(\ref{eq:gphnoad1}) reads
\begin{equation}
\gamma_{+}^{d}=2\pi n -\frac{\beta}{\lambda}\ln(\cos^{2}\frac{\theta_{n}}{2}+
e^{-2\pi\frac{\gamma}{\beta}}\sin^{2}\frac{\theta_{n}}{2}),
\end{equation}
As expected, the decoherence-free case is recovered for low values
of the parameter $\lambda$. In fact in the limit of
$\lambda<<\beta$ the geometric phase results:
\begin{equation}\label{eq:1ordnoad}
\gamma_{+}^{d}= \gamma_{+} + \frac{\lambda}{2\beta}(\pi\sin\theta_n)^2+o((\frac{\lambda}{\beta})^2).
\end{equation}
This result can be easily understood in the following way. In the case of no-jump evolution, i.e. of no photon
emission, the decay rate due to the imaginary part of the Hamiltonian $\tilde{H}$ is proportional to the number
$n$ of photons contained in the mode. Therefore, given an initial superposition of two states with different
values of $n$, the amplitude associated to the lowest number of photons gradually increases in time and eventually
the system converges to the state with the lowest $n$. In other words, since, for the no-jump trajectory, no
photon decaying is observed, the probability of the lower $n$ state increases with time. In the Bloch sphere
representation of the subspace $\{\ket{e,n},\ket{g,n+1}\}$, the evolution under the free
Hamiltonian~(\ref{eq:rotation}) would just appear as a rotation of the Bloch vector around the $z$ axis. By
considering the decoherence of the field the state will then spiral towards the south pole. Thus, the first order
term in $\lambda/\beta$ appearing in Eq.~(\ref{eq:1ordnoad}) accounts for the extra area spanned by the system on
the Bloch sphere due to the decoherence.

As expected the geometric phase is affected by decoherence in
different ways for the adiabatic and non-adiabatic scenarios. In
particular, for low decoherence rates, i.e. $\lambda << R$ and
$\lambda << \beta$ in the adiabatic and non-adiabatic case,
respectively, the lowest correction is quadratic in the former and
linear in the latter case. This should be expected since in the
adiabatic evolution the probability for the state to be dragged
away by the decoherence from the unperturbed evolution is washed
away (in the first order) by the driving Hamiltonian, thereby
opposing against decohering effects. On the other hand, in the
non-adiabatic evolution, there is no action other than the
decoherence, which finds no resistance in the evolution.
%and for an initial state $|e\alpha\rangle$, where $|\alpha\rangle$
%is a coherent state of the field, we find
%\begin{equation}
%\gamma_{\alpha}^{d}=-\frac{\beta}{\lambda}(e^{-2\pi\frac{\lambda}{\beta}}-1).
%\end{equation}

Analogous considerations can be done in the case of a two modes
system described by the Hamiltonian~(\ref{eq:ham2mode}). The
non-adiabatic version of the same problem can be described in the
following way. We assume that we can prepare the system in the
initial state:
\begin{equation}
  \label{eq:instate}
  |\psi_{in}\rangle =e^{-iJ_y\alpha}|\phi\rangle.
\end{equation}
The dynamics of the system is described, in the interaction
picture, by the Hamiltonian:
\begin{equation}
  \label{eq:rotation}
  H_{int}=\delta J_z,
\end{equation}
where $\delta$ is now the constant parameter. Thus, the state
evolves according to:
\begin{equation}
  \label{eq:evol}
  |\psi(t)\rangle=e^{-i\delta J_z t } |\psi_{in} \rangle,
\end{equation}
and as in the one mode case, after a time $T=4\pi/\delta$ the
state completes a closed loop. The non-adiabatic phase is
$\chi_{(n,n^{\prime})}$. Adding decoherence to the system we
obtain the following phase
\begin{eqnarray}
&&\chi_{(n,n^{\prime})}=4\pi(1-\cos\alpha)\frac{n-n^\prime}{2}-\\
&&-1/4\cos\alpha\frac{\delta}{\lambda}
\ln\left[1-\frac{1}{2}(1-\cos{\theta_n})\left(1-e^{-\frac{\lambda}{\delta}8\pi}\right)\right]\nonumber
\end{eqnarray}
and at first order in $\lambda/\delta$ we recover the expression
(\ref{eq:gp2mode}), with $\Omega=4\pi(\cos\alpha)$, i.e. the solid
angle spanned on the parameter sphere in the case of no
decoherence.

Working towards having a realistic description of geometric phases we have introduced field decoherence in the
problem of a two-level system interaction with a quantized field. We analyzed the one and two mode models in the
adiabatic and non-adiabatic case. We showed that when the geometric phase is generated by an adiabatic evolution
the first correction due to the decoherence of the driving field for the no-jump trajectory is only of second
order in the decaying rate of the field $\lambda$. This result reinforces the idea that geometric phases can be
robust to errors, in agreement with previous works which analyze the geometric phase under {\it classical} noise
sources~\cite{gabriele} like random fluctuations of a classical driving field. We also showed that, for the
non-adiabatic evolution this is no longer the case, and decoherence effects appear already in the first order
correction term. This result is also in accordance with previous works in which, again, different {\it classical}
noise sources were considered~\cite{non-adiabatic123,nazir}.

Our results are particularly relevant in the experimental
realizations of these phases, like the one proposed
in~\cite{carollo02}, and in their use in the implementation of
geometric quantum computation. Understanding the effects of
decoherence in the geometric evolution of states is the first step
in finding schemes resilient to this. We are now investigating a
robust scheme to field decoherence using engineered reservoirs.

This research was supported by EPSRC, Hewlett-Packard, Elsag spa and the EU and Quiprocone Grant No. 040. AC
acknowledges the support of "Fondazione Angelo della Riccia" and the University of Milan and MFS acknowledges the
support of CNPq.

%An experimental proposal to measure the fully quantized version of
%the phase has been proposed \cite{carollo02} and the results
%presented here are very relevant to evaluate the feasibility of
%this experiment.

\bibliographystyle{nature}
\bibliography{ivettenew}

\end{document}